\documentclass[aps,pre,twocolumn,groupedaddress]{revtex4-1}
\usepackage{graphicx}
\usepackage{color}
\usepackage{epsfig}
\usepackage{latexsym}
\usepackage{bm}
\usepackage{ulem}
\usepackage{color}

\usepackage{dcolumn}
\usepackage{amsmath}    
\usepackage{amssymb}
\usepackage{bm} 
\usepackage{hyperref}
\usepackage{latexsym}
\usepackage{color}
\usepackage{subfigure}
\def\beq{\begin{equation}}
\def\eeq{\end{equation}}

\def\lsim{\:\raisebox{-0.5ex}{$\stackrel{\textstyle<}{\sim}$}\:}
\def\gsim{\:\raisebox{-0.5ex}{$\stackrel{\textstyle>}{\sim}$}\:}
\def\beq{\begin{equation}}                           
\def\eeq{\end{equation}}                           
\def\bea{\begin{eqnarray}}                           
\def\eea{\end{eqnarray}}        

\begin{document}



\title{Enhanced dynamics of active Brownian particles in periodic obstacle arrays and corrugated channels}

\author{Sudipta Pattanayak}
\email{sudipta.pattanayak@bose.res.in}
\affiliation{S. N. Bose National Centre for Basic Sciences, J D Block, Sector III, Salt Lake City, Kolkata 700106}
\author{Rakesh Das}
\email{rakesh.das@bose.res.in}
\affiliation{S. N. Bose National Centre for Basic Sciences, J D Block, Sector III, Salt Lake City, Kolkata 700106}
\author{Manoranjan Kumar}
\email{manoranjan.kumar@boson.bose.res.in}
\affiliation{S. N. Bose National Centre for Basic Sciences, J D Block, Sector III, Salt Lake City, Kolkata 700106}
\author{Shradha Mishra}
\email{smishra.phy@itbhu.ac.in}
\affiliation{Department of Physics, Indian Institute of Technology (BHU), Varanasi, India 221005}


\begin{abstract}
We study the motion of an active Brownian particle (ABP) using overdamped Langevin dynamics on a two-dimensional substrate with 
periodic array of obstacles and in a quasi-one-dimensional corrugated channel comprised of periodically arrayed obstacles. The periodic 
arrangement of the obstacles enhances 
the persistent motion of the ABP in comparison to its motion in the free space. Persistent motion increases with the activity 
of the ABP. We note that the periodic arrangement induces directionality in ABP motion at late time, and it increases with 
the size of the obstacles. We also note that the ABP exhibits a super-diffusive dynamics in the corrugated channel. 
The transport property is independent of the shape of the channel; rather it depends on the packing fraction of the obstacles 
in the system. However, the ABP shows the usual diffusive dynamics in the quasi-one-dimensional channel with flat boundary.  

\end{abstract}
\maketitle


\section{Introduction \label{Introduction}}

Active systems \cite{ramaswamy, marchetti, romanczuk, bechinger} have been a frontier area of research in the last two decades because of their unusual 
properties as compared to the systems in thermal equilibrium. Natural systems like motile microorganisms, flock of birds, school of 
fishes, etc. and artificial (Janus) micro-particles are some examples of the active systems. The individual constituents 
of these systems transduce their internal energy into motion, i.e., they exhibit self-propulsion characteristics, and therefore, they are 
also called self-propelled particles (SPPs). In addition to the extensive study of these systems in clean environments 
\cite{ramaswamy, marchetti, romanczuk, vicsek1995, tonertu, chate2007, chate2008}, recently people have started to look for their bulk properties in heterogeneous medium 
\textcolor{red} {\cite{quint,morin,chepizhko,rakesh2018,toner2018,chepizhko2013,chepizhko2018, reichhardt1, reichhardt2}}. 

Active Brownian particles (ABPs) \cite{athermalmarchetti} are one kind of SPPs where the particles do not have any mutual alignment interaction, 
and they exhibit many interesting phenomena like motility induced phase separation \cite{cates1, cates2, cates3, athermalmarchetti}. In a recent study, Reichhardt {\it et al.} examine a two-dimensional system of run-and-tumble active matter disks that can exhibit motility induced phase separation interacting with a periodic quasi-one-dimensional traveling-wave substrate. Authors note that the collective clustering of run-and-tumble disks could be
an effective method for forming an emergent object that can
move against gradients or drifts even when individual disks on
average move with the drift \cite{reichhardt3}. In another study, Reichhardt {\it et al.} consider ballistic active disks driven through a random obstacle array. Formation of a pinned
or clogged state occurs at much lower obstacle densities for the active disks than for passive disks \cite{reichhardt4}. Very recently, 
the dynamics of the ABP is shown to be sub-diffusive in the presence of obstacles modeled as a random Lorentz gas for the density of obstacles close 
to the percolation threshold \cite{stark}. These ABPs are shown to attain their long-time dynamics faster than the passive (Brownian) particles, 
because of their persistent motion \cite{stark}. 
In contrast, when obstacles are arranged periodically, it is found that the  persistent length of the active particle increases  \cite{chinnasamy}. Choudhury {\it et al.}, consider chemically boosted self-propelled Janus colloids moving atop a two-dimensional crystalline surface. The authors find that the dynamics of the self-propelled
colloid reflects a competition between hindered diffusion due to the periodic surface and enhanced diffusion due to active motion \cite{udit}. Hence, the nature of the heterogeneous environment modifies the dynamics of the active particles. 

The dynamics of the active particle is not only modified
in heterogeneous substrates, but it can also be modified us-
ing a confined channel. The boundary of the confined wall plays
an important role in the motion of active particles \cite{shradha, moumita, stark3, zhong}. Recently, Dey {\it et al.}, showed that the confinement can enhance the average rate of binding of the motor-cargo complexes to the microtubule, which leads to an enhancement in the average velocity \cite{moumita}. Also the asymmetric channel corrugation induces a net-flux in the motion of microswimmers along the channel, the strength and direction of which strongly depends on the swimmer type \cite{stark3}. Furthermore, a non-zero average drift can be induced in ABP using potential modulation between two directions in a 2D periodic corrugated channel \cite{zhong}.

Motivated by the fact that the arrangement of the obstacles, and a different kind of confined channel can modify the dynamics of the active particles, in the present work we ask the question:
 How does the dynamics of the active particle varies with its activity, and density of the obstacles arranged periodically (i) on a two-dimensional substrate and (ii) along the boundary of a quasi-one-dimensional channel?

To answer the first question, we numerically study the dynamics of an 
ABP on a 2D substrate with periodically arranged obstacles. The ABP shows a cross-over from its initial super-diffusive to diffusive dynamics, and such a cross-over is an intrinsic feature of the active particles \cite{wu,patteson}. We find that, due to the steric interaction between the 
ABP and obstacles, the cross-over time of the ABP increases with its self-propulsion speed.  Furthermore, we note that in a dense obstacle environment the ABP performs a more directed motion. In the later part of this paper, the dynamics of the ABP in a quasi-one-dimensional corrugated channel comprised of periodically arrayed obstacles is studied. We find that the corrugated channel governs a super-diffusive dynamics of the ABP along the channel without any external drive. Also the transport is independent of the shape of the corrugated boundary, and it only depends on the packing fraction of the obstacles in the channel. However, we find the flat boundary does not encourage the super-diffusive motion. 

The rest of the article is organized as follows. In section \ref{Model} we introduce the microscopic rule based model for the ABP in periodic 
geometries. The results of the numerical simulation of a 2D substrate with periodic obstacles and corrugated channel are given in section \ref{periodic substrate} and \ref{corrugated channel}, respectively. Finally in section \ref{discussion}, we discuss our results and future prospect of our study.



\section{Model \label{Model}}
\begin{figure}[t]
\centering
\includegraphics[width=1.0\linewidth]{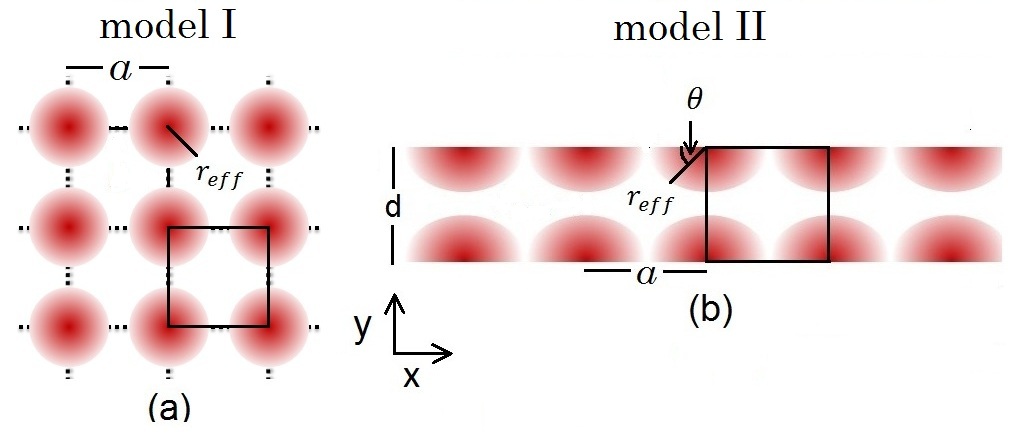}
\caption{(Color online) : (a) The schematic picture of a 
square lattice with obstacles at its vertices. Centre to centre distance between obstacles $a = 1.0$. The packing fraction of the lattice is varied from $\Phi = 0.125$ (obstacle free substrate) to $\Phi = 0.39$. (b) The schematic picture of a quasi-one-dimensional corrugated channel comprised periodically arrayed circular / elliptical obstacles. The periodicity $a$, and width of the channel $d$ are shown. $r_{eff}$ (defined in the text) is shown. The $\Phi$ of the channel is varied from $\Phi = 0.10$ to $0.60$ by changing d or $a$. Boxes show unit cell for both cases. $x$ and $y$  directions for both model are shown.}
\label{figure_square}
\end{figure}

We consider a circular-disk shaped active Brownian particle of radius $R_p$ placed in a periodic obstacle environment. Its dynamics is 
studied for two models; (i) in model I, we consider a 2D $L \times L$ square lattice, where circular-disk shaped obstacles of radius 
$R_{o}$ are placed periodically at the vertices, and (ii) in model II, we consider a quasi-one-dimensional corrugated channel comprised 
periodically arranged circular or elliptical obstacles at the boundary of the channel. The semi-major and the semi-minor axes of the 
elliptical obstacles are designated by $max(a^{\prime}, b^{\prime})$ and $min(a^{\prime}, b^{\prime})$, respectively. $a^{\prime}$ and 
$b^{\prime}$ are always chosen along the $x$ and $y$-axes, respectively, as shown in Fig.~\ref{figure_square}. For a corrugated channel 
with circular obstacles $a^{\prime} = b^{\prime}$. Let us represent the position vector of the centre of the ABP by ${\bm r}(t)$ at 
time $t$. The ABP moves along its orientation defined by a unit vector ${\bm e}(t)$ in the $x$-$y$ plane. The dynamics of the ABP is 
governed by the overdamped Langevin equation 
\begin{eqnarray}
\dfrac{d{\bm r}(t)}{dt} &=& v_{0}{\bm e}(t) + \mu \sum_{i} {\bm F}_{0}^{i}, \label{equation_position} \\
\dfrac{d{\bm e}(t)}{dt} &=& \sqrt{2D^{R}}{\bm \eta}^{R}(t) \times {\bm e}(t), \label{equation_direction}
\end{eqnarray}
The first term on the right-hand-side (RHS) of Eq.~\ref{equation_position} is due to the activity of the ABP, and its self-propulsion speed is $v_{0}$. The second term represents the
steric force acting on the ABP due to its neighboring obstacles, and it is tuned by a parameter $\mu$, which is $0$ for the obstacle-free 
substrate and $1$ for all other cases. We consider ${\bm F}_0 = -\nabla V$, where the steric interaction is incorporated by the 
{\it Weeks-Chandler-Anderson} potential defined as
\begin{eqnarray}
 V &&= 4\epsilon \left[ \left(\frac{\sigma}{|{\bm r}-{\bm r}_o|}\right)^{12} - \left(\frac{\sigma}{|{\bm r}-{\bm r}_o|}\right)^{6} \right] 
				+ \epsilon, \notag \\ 
   &&\quad\quad\quad\quad\quad\quad\quad\quad\quad\quad\quad {\textrm {for}} \quad |{\bm r}-{\bm r}_o| < r_{eff}, \notag \\
   &&= 0, \quad\quad\quad\quad\quad\quad\quad\quad\quad {\textrm {for}} \quad |{\bm r}-{\bm r}_o| \geq r_{eff}.
\end{eqnarray}
Here ${\bm r}_o$ represents the position vector of the centre of a neighboring obstacle. We treat the ABP as a point point particle in 
our simulation, and its size is taken care by an effective radius $r_{eff}$ of the obstacles. While in model I, $r_{eff} = R_{p} + R_{o}$, 
in model II, $r_{eff} = R_p + a^{\prime}b^{\prime} / \sqrt{a^{\prime 2}\sin^2\theta + b^{\prime 2}\cos^2\theta}$, where $\theta$ is 
the angle of ${\bm r} - {\bm r}_o$ with respect to the $x$-axis.
We consider $\epsilon = 1$, and the parameter $\sigma = r_{eff} / (2^{1/6})$.


The rate of change of the orientation ${\bm e}(t)$ of the ABP is given by Eq.~\ref{equation_direction}. $D^{R}$ represents the rotational 
diffusion constant, and ${\bm \eta}^{R} = \eta_{z}^{R} {\bm e}_{z}$ is the stochastic torque with zero mean and Gaussian white noise 
correlations, {\it i.e.},
\begin{eqnarray}
<{\bm \eta}^{R}(t)> &=& 0, \\
<{\bm \eta}^{R}(t_{1}) \otimes {\bm \eta}^{R}(t_{2})> &=& {\bf 1} \delta (t_{1} - t_{2}).
\label{equation_noise}
\end{eqnarray}
Note that the stochastic torque always points out of the substrate, {\it i.e.}, along the $z$-axis.

The schematic of the models I and II are shown in Fig.~\ref{figure_square}(a) and (b), respectively, and the closed boxes represents the respective 
unit cells. Colors in Fig. \ref{figure_square} shows the intensity plot of the potential. White regions are zero-potential regions, and the repulsive 
potential increases from white to dark red. Fig.~\ref{figure_square}(a) depicts a square lattice with spacing $a = 1$. 
We define the packing fraction $\Phi$ of the system as the fraction of the area of a unit cell occupied by the obstacles and the ABP. Therefore, 
in model I the packing fraction is given by $\Phi = (\pi {R_{o}}^{2} + \pi {R_{p}}^{2}) / a^{2}$. We vary $\Phi$ from $0.125$ (obstacle free substrate) 
to $0.39$ by changing $R_o$ so that the ABP does not get confined in a unit cell and it can pass through the obstacles.

In Fig.~\ref{figure_square}(b) a corrugated channel of width $d$ ({\it i.e.}, centre to centre separation of two neighboring obstacles 
in the $y$-direction) is shown schematically. The channel is composed of elliptical or circular-disk shaped obstacles arrayed along the 
$x$-direction with periodicity $a$. The packing fraction for the corrugated channel is defined as, 
$\Phi = (\pi a^{\prime} b^{\prime} + \pi {R_{p}}^{2}) / ad$. 
We vary $\Phi$ in model II from $0.10$ to $0.60$. The surface to surface separation of the obstacles are chosen such that the ABP cannot 
pass through the obstacles the $y$-direction.

The dimensionless angular Peclet number is defined as $Pe = v_{0} / D_{R} R_{p}$.  The persistent length of the particle is defined as 
$l = v_{0}/D_{R}$, and the corresponding persistent time $\tau = 1/D_{R}$. The rotational diffusion constant is kept fixed at $D_{R} = 0.1$, 
and $v_{0}$ is varied in our study. 
Initially the ABP is placed randomly in one of the unit cells with random ${\bm e}$. The dynamics of the ABP is studied using the evolution 
Eqs.~(\ref{equation_position})-(\ref{equation_direction}). Periodic boundary condition is used in both directions for model I and in 
$x$-direction for model II. Simulation is done for total time steps $10^6$ and $10^7$ for models I and II, respectively, and the smallest time step considered is $\Delta t = 10^{-3}$. All the physical quantities calculated here are averaged over $10000$ realizations.

\section{Results \label{results}}
\subsection{Substrate with periodic array of obstacles \label{periodic substrate}}


\begin{figure}[t]
\centering
\includegraphics[width=1.0\linewidth]{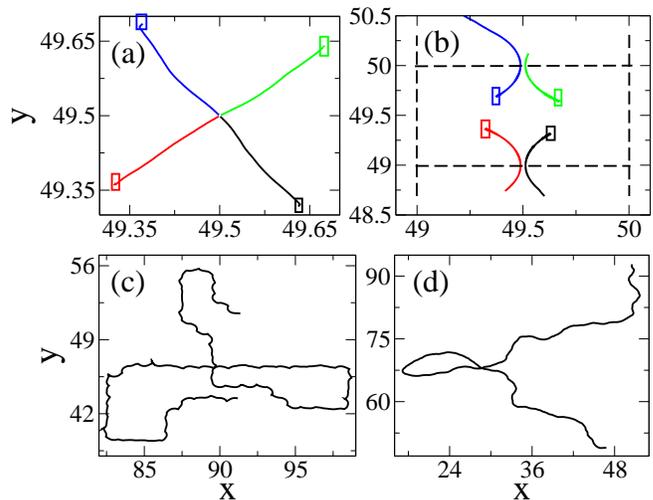}
\caption{(Color online) The plot of the ballistic trajectories of four ABPs at the beginning and when they follow the obstacle boundary are shown in (a) and (b) respectively. The initial coordinate for all ABPs is $(49.5,49.5)$, 
and their directions are different. Four different colors are used for the four ABPs. The intersection points of the dashed lines in (b) represent the centre of an obstacle. The boxes in (a) represent the end point of the trajectories, and boxes in (b) represent the starting of the trajectories. $\Phi = 0.39$. The plot of late time diffusive trajectory of an ABP on the two dimensional periodic obstacle substrate of $\Phi = 0.39$, and  $\Phi = 0.125$ (free substrate) is shown in (c) and (d) respectively. The time interval is the same $(100)$ in (c) and (d). We consider $Pe = 50$.}
\label{figure_square_trajectory}
\end{figure} 

We first study the dynamics of the ABP on a 2D substrate with periodic array of obstacles, {\it i.e.}, for model I. Typical trajectories
of the ABP are shown in Fig.~\ref{figure_square_trajectory}. The ballistic motion at the beginning for four ABPs with different initial 
direction is shown in Fig.~\ref{figure_square_trajectory}(a). The interplay of obstacle hindrance and $D_{R}$ causes the ABP to follow the obstacle boundary, which is shown in Fig.~\ref{figure_square_trajectory}(b). This phenomena are also present at late time motion. The late 
time trajectories of the ABP on the 2D substrate with periodic obstacles and in the obstacle free space are shown in Fig.~\ref{figure_square_trajectory}(c) 
and (d) respectively. An interesting point to note from these two figures is that the late time trajectory of the ABP shows a more directional motion in a periodic obstacle environment in comparison to the free space. 

\begin{figure}[t]
\centering
\includegraphics[width=1.0\linewidth]{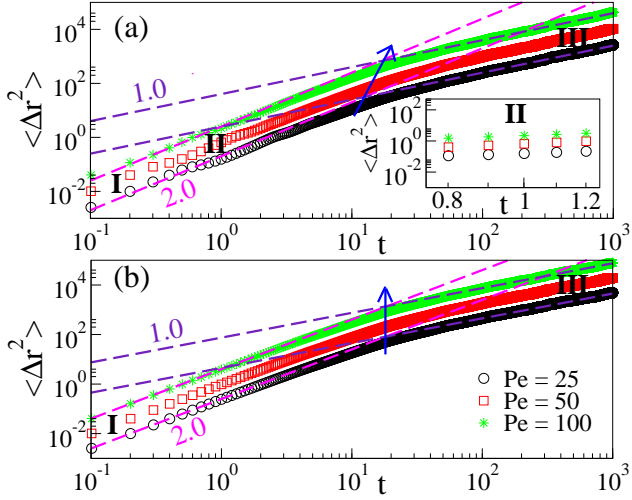}
\caption{(Color online) Plot of the mean square displacement of the ABP $\left\langle \Delta r^{2} \right\rangle $ vs. time $t$ in the periodic square lattice of $\Phi = 0.39$ (a) and $\Phi = 0.125$ (obstacle free substrate)(b). Region I and III are the ballistic and diffusive regions of the ABP respectively. The lines of slope $2$ (magenta) and $1$  (indigo) are shown. The approximate cross-over points from super-diffusive to diffusive dynamics for different $Pe$ for both cases are shown by a blue arrow. In inset of (a): $\left\langle \Delta r^{2} \right\rangle $ with time $t$ for different $Pe$ in region II (when ABP moves along obstacle boundary) is shown.}
\label{figure_msd_square_lattice}
\end{figure}


\begin{figure}[t]
\centering
\includegraphics[width=1.0\linewidth]{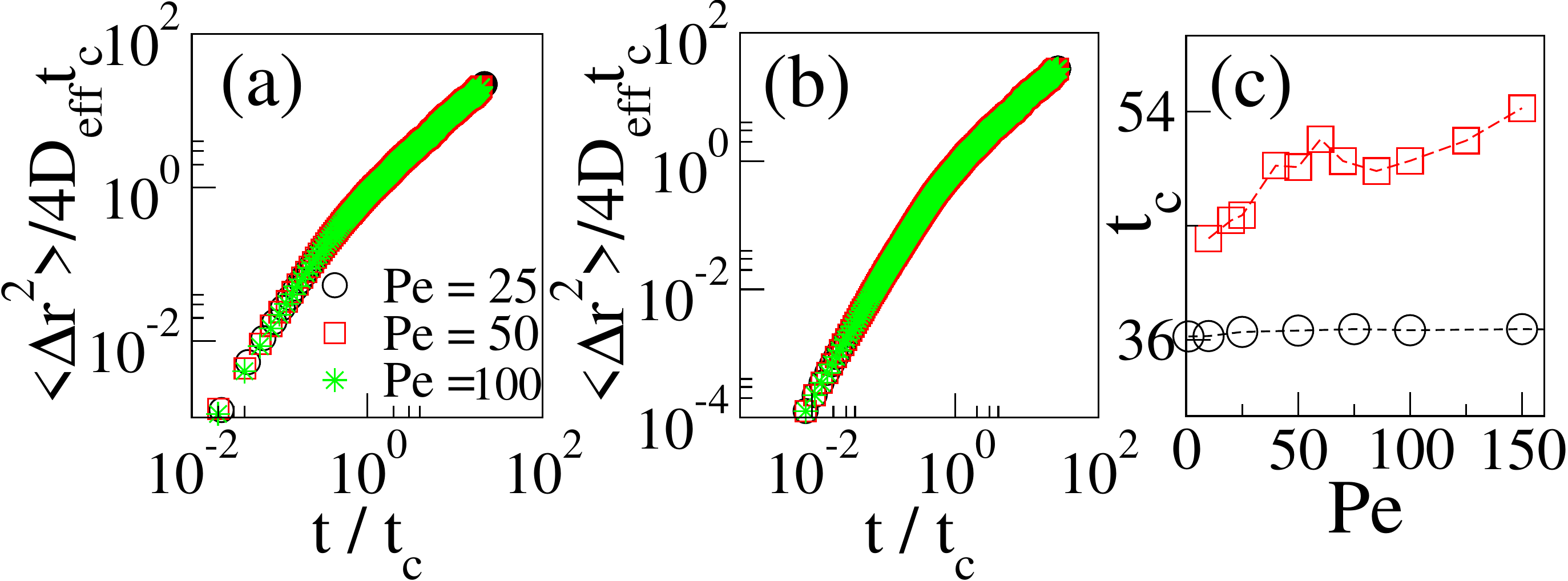}
\caption{(Color online) Plot of the scaled mean square displacement $\left\langle \Delta r^{2} \right\rangle / 4D_{eff}t_{c}$ vs. scaled time $t/t_{c}$ of the ABP in the square lattice of $\Phi = 0.39$ (a) and $\Phi = 0.125$ (obstacle free substrate) (b) is shown. (c) The cross-over time $t_{c}$ 
with $Pe$ for $\Phi = 0.39$ and $\Phi = 0.125$(obstacle free substrate) are shown by red square boxes and black circles respectively.}
\label{figure_msd_square_scaling}
\end{figure}


\begin{figure}[t]
\centering
\includegraphics[width=1.0\linewidth]{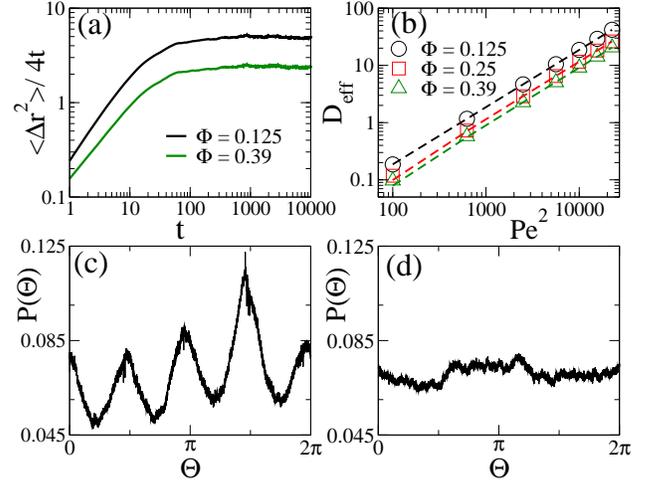}
\caption{(a) Variation of $<\Delta r^{2}>/4t$ with time $t$ for $Pe = 50$. The black and green lines are for $\Phi = 0.125$(obstacle free substrate) and $\Phi = 0.39$, respectively. (b) Plot of the effective 
translational diffusion constant $D_{eff}$ of the ABP for different $Pe$.
The black circles, red squares and blue triangles are for the periodic $\Phi = 0.125$ (obstacle free substrate), $\Phi = 0.25$ and $\Phi = 0.39$ respectively. The linear slopes for $\Phi = 0.125, 0.25$ and $0.39$ are 0.0018, 0.0011, 0.0009, respectively. The plot of  $P(\Theta)$ of the ABP for $\Phi = 0.39$ and $\Phi = 0.125$ (obstacle free substrate) is shown in (c) and (d) respectively. For (c) and (d) we consider $Pe = 50$.}
\label{figure_square_diff_theta_prob} 
\end{figure}


To characterize the dynamics of the ABP, we calculate its mean square displacement (MSD) defined as 
\begin{eqnarray}
<\Delta && r^2 (t)> = \notag \\
&&\frac{1}{N} \sum_{n=1}^{N} \left[\left(x_{n}(t)-x_{n}(0)\right)^{2}+\left(y_{n}(t)-y_{n}(0)\right)^{2}\right],
\label{msd_equation}
\end{eqnarray}
where $N$ is the total number of realizations, $x_{n}(t)$ and $y_{n}(t)$ represent the respective coordinates 
of the ABP at time $t$ for the $n^{th}$ ensemble in the $x$-$y$ plane. The MSD of the ABP in a periodic obstacle environment and in a free 
substrate for different $Pe$ are shown in Fig.~\ref{figure_msd_square_lattice}(a) and (b), respectively. The ABP performs a persistent random 
walk, which is one of the common features in the active systems \cite{stark, chinnasamy, udit, wu}. Therefore, the MSD can be written as,
\begin{equation}
<\Delta r^{2}> = 2 \mathfrak{D} D_{eff} t \left[1 - exp\left(-\frac{t}{t_{c}}\right)\right],
\label{equation_deff}
\end{equation}
where $\mathfrak{D}$ represents dimensionality of the space, $D_{eff}$ is the effective diffusion constant in the steady state, and $t_c$ 
is the cross-over time from the initial ballistic regime $<\Delta r^{2}> = 4 D_{eff} t^2/t_c$ for $t<<t_c$ to the late time diffusive regime 
$<\Delta r^{2}> = 4 D_{eff} t$ for $t>>t_c$.
The two lines of slope $2$ and $1$ shown in Fig.~\ref{figure_msd_square_lattice} represent the ballistic (I) and the diffusive (III) regimes 
of the ABP, respectively, for different $Pe$. We estimate the effective diffusivity $D_{eff}$ from the asymptotic limit of $<\Delta r^{2}>/4t$ vs. $t$ variation as shown in Fig. \ref{figure_square_diff_theta_prob} (a), and the cross-over time $t_{c}$ is estimated by fitting numerical data with Eq. \ref{equation_deff}. The cross-over time $t_c$ for the obstacle free environment does not change with $Pe$, but $t_c$ changes with $Pe$ for the 
periodic obstacle environment. The approximate change in $t_{c}$ is shown by arrows in Fig.~\ref{figure_msd_square_lattice} (a) and (b). 
The ABP also realizes a small confinement effect (regime with label II) in the presence of the obstacles during its persistent motion,
and MSD displays a plateau for that time duration, which is shown in the inset of Fig.~\ref{figure_msd_square_lattice}(a); this kind of confinement is also present at long time. The scaled MSD 
$<\Delta r^2>/4 D_{eff} t_c$ versus scaled time $t/t_c$ for different $Pe$ for the periodic obstacle and the obstacle free environment is plotted 
in Fig.~\ref{figure_msd_square_scaling}(a) and (b), respectively. In both the cases data shows a good scaling collapse. The plot of $t_{c}$ 
vs. $Pe$ for the periodic obstacle (squares) and obstacle free substrate (circles) is shown in Fig.~\ref{figure_msd_square_scaling}(c). 
The $t_{c}$ changes with $Pe$ for the periodic obstacle substrate, whereas it is constant for the free substrate. Also $t_c$ is larger for the 
periodic obstacle environment as compared to the free case. Therefore the periodicity enhances the persistence motion of the ABP. 

The variation in the effective diffusion constant $D_{eff}$ with $Pe^{2}$ 
for different $\Phi$ is shown in Fig.~\ref{figure_square_diff_theta_prob}(b). The enhanced diffusion is one of the intrinsic feature in the active 
systems, as found before in \cite{aparna2}. We find that the effective diffusivity $D_{eff}$ of the ABP for a fixed $v_{0}$ decreases as we increase $\Phi$. For $\Phi = 0.39$ and $0.125$, $D_{eff} \sim Pe^2$ with slope $0.0009$ and $0.0018$ respectively. Interestingly, $D_{eff}$ in the dense periodic ($p$) obstacle environment is exactly half of its value in the obstacle
free ($f$) space. In the steady state, the MSD of the ABP can be expressed by $2 \mathfrak{D}_{eff} D_{eff}^{p/f} t$, where $\mathfrak{D}_{eff}$ 
is the effective dimensionality of the space and $D_{eff}^{p/f}$ represents the effective diffusivity in the periodic obstacle / obstacle free environment. 
Since $D_{eff}^p = \frac{1}{2} D_{eff}^f$, that implies the effective dimensionality for the system for dense periodic array of obstacles reduces to one.
To further explain this, we calculate the probability distribution function $P(\Theta)$ of the instantaneous orientation $\Theta$ of the ABP in the 
steady state. The plot of $P(\Theta)$ for the periodic obstacle and the obstacle free substrate is shown in Fig.~\ref{figure_square_diff_theta_prob}(c) 
and (d), respectively. $P(\Theta)$ shows peaks for the dense periodic obstacle environment and the magnitude of one peak is always larger. The height of the peaks decreases as we decrease $\Phi$ (data is not shown). However, $P(\Theta)$ becomes flat for the obstacle  free environment. Therefore the ABP moving in a dense periodic obstacle environment shows a directional preference during its motion. This explains why the $D_{eff}$ of the ABP in a periodic environment is half of its value in free space.

The periodic arrangement of the obstacles enhances the persistent motion of the ABP, and at late time, the motion is more like 
a one-dimensional persistent random walk. This phenomenon of the ABP is not present either in random obstacles \cite{stark} or in free environment. The immediate question arises as to what will happen if we restrict the motion of the ABP along one direction only. In the next part of this paper we study the dynamics of the ABP in a quasi-one-dimensional corrugated channel as shown in Fig.~\ref{figure_square}(b).

\subsection{Corrugated channel \label{corrugated channel}}

\begin{figure}[t]
\centering
\includegraphics[width=1.0\linewidth]{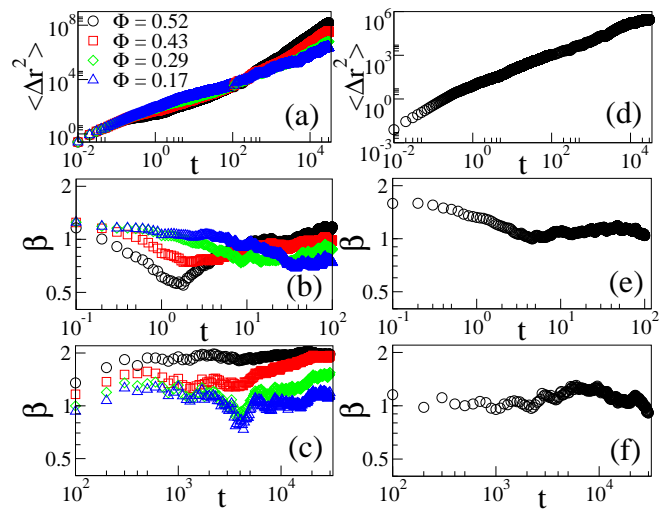}
\caption{(Color online) Plots of the mean square displacement
 $\left\langle \Delta r^{2} \right\rangle$ , the exponent $\beta$ at early and late time of the ABP in the corrugated channel for different $\Phi$ are shown in (a-c), respectively. We consider $Pe = 100$ and $\Phi$ changes as we vary the channel width d. $\left\langle \Delta r^{2} \right\rangle$, the exponent $\beta$ at early and late time of the ABP in a flat repulsive channel of width $d = 0.42$ are shown in (d-f), respectively. For the flat channel the radius of the ABP is $r_p=0.2$, and $Pe = 50$.}
\label{figure_corrugated_msd_diff_width}
\end{figure}

First we consider a corrugated channel comprised of circular-disk shaped obstacles with periodicity $a$ and width $d$. The radii of each obstacle 
and the ABP are chosen as $R_o = 0.29$ and $R_p=0.2$, respectively. 
The dynamics of the ABP is characterized by its MSD as defined in Eq.~(\ref{msd_equation}) and a MSD exponent $\beta$ such that 
$\left\langle \Delta r^{2}(t) \right\rangle \sim t^{\beta}$. This exponent $\beta$ can also be defined as
\begin{equation}
\beta(t) = log_{10} \dfrac{\left\langle \Delta r^{2}(10t) \right\rangle}{\left\langle \Delta r^{2}(t) \right\rangle}.
\end{equation}
The exponent $\beta = 2$ and $1$ for the ballistic and the diffusive dynamics, respectively. We fixed the periodicity of the channel and changed the width of the channel to vary the $\Phi$ of the system. The  MSD for different $\Phi$ is shown in Fig.~\ref{figure_corrugated_msd_diff_width}(a), and we calculate the $\beta$ from MSD data. We note that at early 
time $t \lsim 100$, the exponent 
$\beta<1$  for large $\Phi$, {\it i.e.}, the ABP exhibits sub-diffusive dynamics for high packing fraction and it exhibits diffusive dynamics ($\beta=1$) for low packing fraction, as shown in Fig.~\ref{figure_corrugated_msd_diff_width}(b). 
However, at late time $t\gsim 100$, the ABP shows a super-diffusive behavior ($\beta>1$) only for a high packing fraction ($\Phi = 0.52, 0.43$), whereas for a low packing fraction ($\Phi = 0.17$) of the channel, the dynamics is diffusive ($\beta=1$), as shown in Fig.~\ref{figure_corrugated_msd_diff_width}(c). Here we consider $Pe = 100$, and we also note a similar behavior for $Pe = 50$. 


\begin{figure}
\centering
\includegraphics[width=1.0\linewidth]{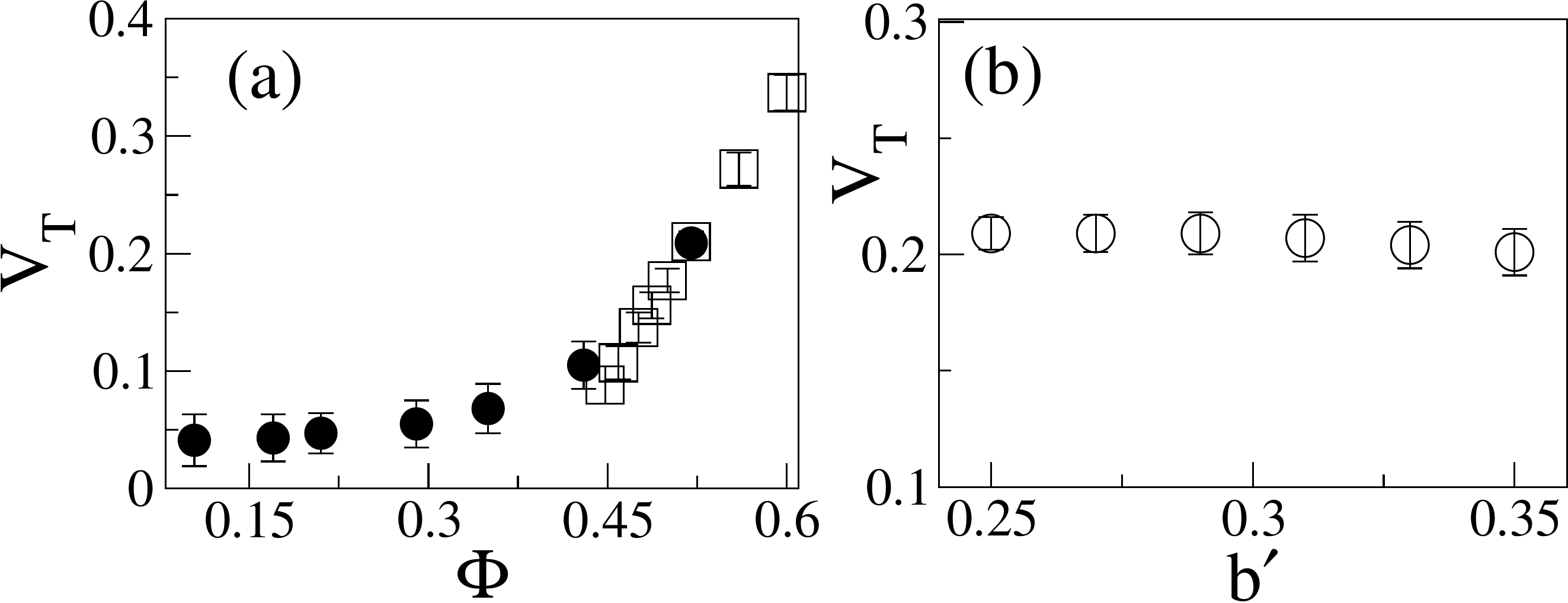}
\caption{(Color online)(a) Plot of the transport speed $V_{T}$ of the ABP in the corrugated channel with packing fraction $\Phi$. We varied $\Phi$ from 0.10 to 0.60. For filled circles, we change the channel width d to vary $\Phi$, and for square boxes, $\Phi$ is changed by varying the periodicity $a$ of the obstacles along the boundary of the channel.  (b) Plot of $V_{T}$ of the ABP in the corrugated channel comprised of periodically arrayed elliptical obstacles vs.  $b^{\prime}$. We fixed the $\Phi = 0.52$ and $a^{\prime} = 0.29$. For (a) and (b) $Pe = 100$. The error bar of $V_{T}$ is shown for all cases.}
\label{figure_corrugated_vt_diff_width}
\end{figure}



To understand the importance of the corrugated geometry, we also calculate the MSD of the ABP in a quasi-one-dimensional channel with flat 
boundary. We note that the ABP performs diffusive motion in the flat geometry, as is evident from Fig.~\ref{figure_corrugated_msd_diff_width}(d-f) drawn for the channel width $d=0.42$ and ABP radius $R_p=0.2$. Therefore, the quasi-one-dimensional corrugated channel drives 
the ABP towards super-diffusive dynamics ($\beta>1$) for sufficiently large time as shown in Fig. 6(c). But after very long time ($ \sim 3 \times 10^4$) the ABP changes its direction due to its rotational diffusion, and further moves in opposite direction for a similar period of time. Hence, the periodicity of the corrugated channel leads to
a much larger ($ \sim 10^4$) persistent time / motion of the ABP for a high packing fraction $\Phi$ of the obstacles.  
 
The induced directionality in the quasi-one-dimensional corrugated channel motivates us to look for a net transport of the ABP 
through the channel. The transport is explored through statistical averages, specifically through the absolute value of the mean displacement, $\Delta r(t) = \sqrt{\left\langle \Delta r^{2}(t) \right\rangle}$. 
The transport speed is defined as $V_{T}= \frac{1}{v_{0}} (\Delta r(t)/t)$. 
The $V_{T}$ for different packing fraction $\Phi$ of the obstacles in the channel are shown in Fig.~\ref{figure_corrugated_vt_diff_width}(a). We note that $V_T$ increases with $\Phi$. We can tune the $\Phi$ of the channel either by decreasing the channel width $d$, or by placing the obstacles more periodically (by decreasing $a$). Therefore, a corrugated channel with closely placed circular-disk shaped obstacles speeds up the net transport of the ABP.

To study how the super-diffusive transport of the ABP depends on the shape of the corrugated channel, we consider a quasi-one-dimensional 
corrugated channel comprised of periodically arrayed elliptical-disk shaped obstacles, as described in Sec.~\ref{Model}. The earlier described case of circular-disk shaped obstacles is a special case of the elliptical obstacles when $a^{\prime}=b^{\prime}$. We keep $a^{\prime}=0.29$ fixed, 
and vary $b^{\prime}$ such that shape of the elliptical obstacles changes from oblate to prolate. $b^{\prime}$ has been varied by varying the width $d$ of the channel. $b^{\prime}$ and width $d$ are chosen such that the packing fraction $\Phi$ of the channel remains constant. We note that $V_T$ does not depend on $b^{\prime}$ for a particular value of $\Phi$, as shown in Fig.~\ref{figure_corrugated_vt_diff_width}(b).Therefore, the transport speed of the ABP in a corrugated channel does not depend on  the shape of the corrugated channel.


\section{Discussion \label{discussion}}
In the first part of this paper, we have studied dynamics of an ABP in the presence of circular-disk shaped obstacles arrayed periodically on a 2D substrate. In the presence of the periodically arrayed obstacles, the 
cross-over time from ballistic to diffusive dynamics of the ABP increases with its activity. We find that the induced directionality in ABP motion increases with the packing fraction of the obstacles. The motion of the ABPs is  directional in a crowded environment when obstacles are arrayed in periodic fashion.     

Motivated by the induced directed motion of the ABP in a periodic crowded environment, in the second part of this paper, we have studied the motion of the ABP in a quasi-one-dimensional corrugated channel, where the motion of the ABP is confined along one direction. We find the super-diffusive dynamics of the ABP over a long time in the quasi-one-dimensional corrugated channel without any external drive. This makes our study different from the previous studies, 
where the net transport of the ABP is observed with an asymmetric corrugated channel \cite{stark3} or using potential modulation in a corrugated channel\cite{zhong}. The net transport of the ABP in a corrugated channel does not depend on the shape of the wall. The transport speed only depends on the packing fraction of the obstacles in the system. However, the ABP shows the usual diffusive dynamics in a channel with flat boundary.

Hence the channel with corrugated wall, the activity of the ABP lead to super-diffusive dynamics of the ABP without any external 
drive. Such transport  is useful to understand the
dynamics of biological microorganisms, intercellular particles, since those often encounter a crowded environment during
their motion. This model provides a significant understanding of the dynamics of the self-propelled particles in confined geometry, which can be verified in experiments and may be helpful for designing an efficient
transport mechanism. In our current study, we have ignored
the inter-particle interaction. It is also interesting to study the
dynamics of the interacting ABPs in different kinds of confined
geometries. \\
    
\section{Acknowledgment  \label{acknowledgement}}
SP and SM would like to thank Department of Physics, Indian Institute of Technology (BHU), Varanasi and S. N. Bose National Centre for 
Basic Sciences, respectively, for their kind hospitality. \\ 

\section{Author contribution statement}
SP, RD, MK and SM designed the project. SP and RD developed the numerical code, and SP executed it. All the authors contributed equally in 
analyzing the results and preparing the manuscript.

\appendix
\section{Graphical abstract}
\begin{figure}[b]
\centering
\includegraphics[width=1.0\linewidth]{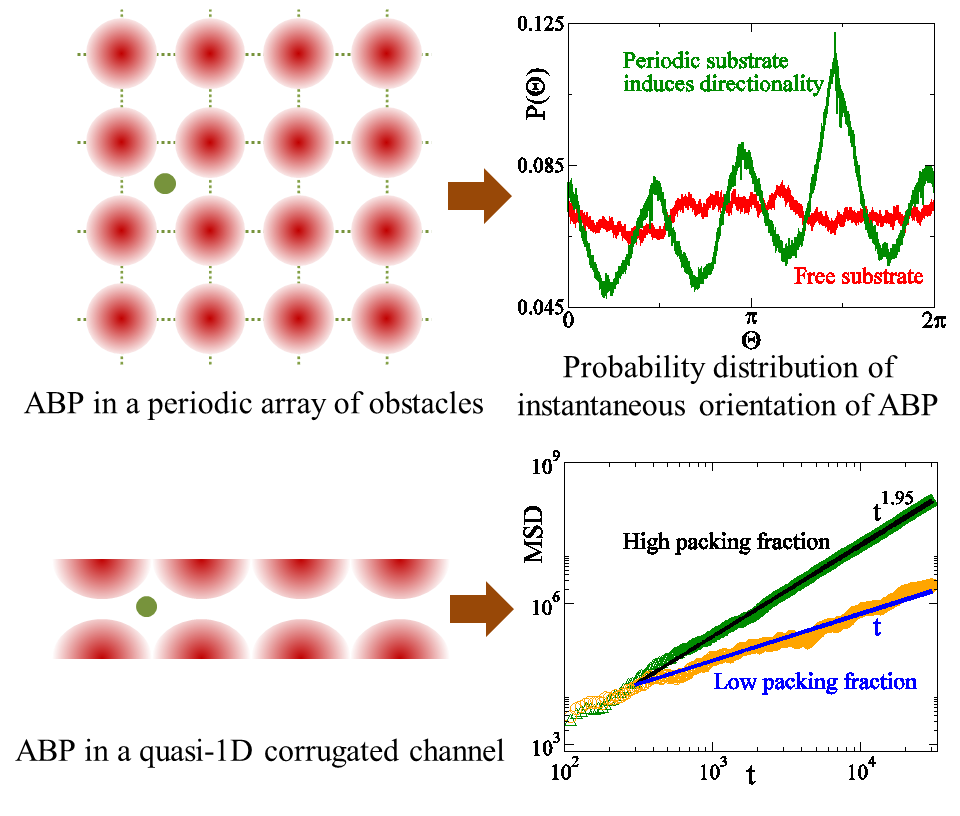}
\end{figure}

\end{document}